\input epsf
\openup 1 pt
\magnification\magstephalf
\overfullrule 0pt
\voffset -1.5truecm
\vsize 24.5 truecm
\def\gsim{\raise.3ex\hbox{$\;>$\kern-.75em\lower1ex\hbox{$\sim$}$\;$}}
\def\P{I\!\!P}
\font\rfont=cmr10 at 10 true pt
\def\ref#1{$^{\hbox{\rfont {[#1]}}}$}

\font\helv=phvr8r scaled 975
\font\bold=phvb8r scaled 975
\font\oblique=phvro8r scaled 975

\font\tenbfit=cmbxti10
\font\sevenbfit=cmbxti10 at 7pt
\font\fivebfit=cmbxti10 at 5pt
\newfam\bfitfam 
\textfont\bfitfam=\tenbfit  \scriptfont\bfitfam=\sevenbfit
\scriptscriptfont\bfitfam=\fivebfit

\font\tenbfit=cmbxti10
\font\sevenbfit=cmbxti10 at 7pt
\font\fivebfit=cmbxti10 at 5pt
\newfam\bfitfam 
\textfont\bfitfam=\tenbfit  \scriptfont\bfitfam=\sevenbfit
\scriptscriptfont\bfitfam=\fivebfit

\font\tenbit=cmmib10
\newfam\bitfam
\textfont\bitfam=\tenbit%

\font\tenmbf=cmbx10
\font\sevenmbf=cmbx7
\font\fivembf=cmbx5
\newfam\mbffam
\textfont\mbffam=\tenmbf \scriptfont\mbffam=\sevenmbf
\scriptscriptfont\mbffam=\fivembf

\font\tenbsy=cmbsy10
\newfam\bsyfam 
\textfont\bsyfam=\tenbsy%


\def\pmb#1{\setbox0=\hbox{#1}
 \kern.05em\copy0\kern-\wd0 \kern-.025em\raise.0433em\box0 }

\def\slash{/\kern-.5em}

\def \half {{\textstyle {1 \over 2}}}

 %


\def\boxit#1{\vbox{\hrule\hbox{\vrule\kern1pt\vbox
{\kern1pt#1\kern1pt}\kern1pt\vrule}\hrule}}

\def\h{\hfill\break}
\parskip=6pt
\parindent=0pt
\hsize=17truecm\hoffset=-5truemm
\def\footnoterule{\kern-3pt
\hrule width 17truecm \kern 2.6pt}


\catcode`\@=11 

\def\nolabels{\def\wrlabeL##1{}\def\eqlabeL##1{}\def\reflabeL##1{}}
\def\writelabels{\def\wrlabeL##1{\leavevmode\vadjust{\rlap{\smash%
{\line{{\escapechar=` \hfill\rlap{\sevenrm\hskip.03in\string##1}}}}}}}%
\def\eqlabeL##1{{\escapechar-1\rlap{\sevenrm\hskip.05in\string##1}}}%
\def\reflabeL##1{\noexpand\llap{\noexpand\sevenrm\string\string\string##1}}}
\nolabels
\global\newcount\refno \global\refno=1
\newwrite\rfile
\def\defref{$^{{\hbox{\rfont [\the\refno]}}}$\nref}
\def\nref#1{\xdef#1{\the\refno}\writedef{#1\leftbracket#1}%
\ifnum\refno=1\immediate\openout\rfile=refs.tmp\fi
\global\advance\refno by1\chardef\wfile=\rfile\immediate
\write\rfile{\noexpand\item{#1\ }\reflabeL{#1\hskip.31in}\pctsign}\findarg}
\def\findarg#1#{\begingroup\obeylines\newlinechar=`\^^M\pass@rg}
{\obeylines\gdef\pass@rg#1{\writ@line\relax #1^^M\hbox{}^^M}%
\gdef\writ@line#1^^M{\expandafter\toks0\expandafter{\striprel@x #1}%
\edef\next{\the\toks0}\ifx\next\em@rk\let\next=\endgroup\else\ifx\next\empty%
\else\immediate\write\wfile{\the\toks0}\fi\let\next=\writ@line\fi\next\relax}}
\def\striprel@x#1{} \def\em@rk{\hbox{}} 
\def\lref{\begingroup\obeylines\lr@f}
\def\lr@f#1#2{\gdef#1{\defref#1{#2}}\endgroup\unskip}
\newwrite\lfile
{\escapechar-1\xdef\pctsign{\string\%}\xdef\leftbracket{\string\{}
\xdef\rightbracket{\string\}}}

\def\writestop{\def\writestoppt{\immediate\write\lfile{\string\p
ageno%
\the\pageno\string\startrefs\leftbracket\the\refno\rightbracket%
\string\def\string\secsym\leftbracket\secsym\rightbracket%
\string\secno\the\secno\string\meqno\the\meqno}\immediate\closeout\lfile}}
\def\writestoppt{}\def\writedef#1{}
\catcode`\@=12 

\helv
\vskip -9truemm\rightline{DAMTP-2019-18}

\centerline{\bold Small \pmb{$t$} elastic scattering and the  \pmb{$\rho$} parameter}
\vskip 3mm
\centerline{A Donnachie}

\centerline{University of Manchester}
\vskip 3mm
\centerline{P V Landshoff}

\centerline{University of Cambridge}
\vskip 2mm

{\bold Abstract}
\helv
A simple application of Regge theory, with 9 free parameters, provides a good 
fit to elastic scattering data at small $t$ from 13.76 GeV to 13 TeV. It
yields a value for $\rho$, the ratio of the real part of the hadronic 
contribution to the forward amplitude to its 
imaginary part, close to 0.14 at 13 TeV. Although the exact value obtained for
$\rho$ is sensitive to what functional form is chosen for the fit,  there is no strong 
case for the presence of an odderon contribution to forward scattering.
\vskip 2 truemm
{\bold 1 The fit}

There are two approaches to the extraction of the phase of the forward elastic scattering amplitude from the data. That favoured by experimentalists is to fit the differential cross section at just one energy beyond the Coulomb region and extrapolate it into the Coulomb region.  This then gives the the hadronic  amplitude up to an unknown phase, which is then determined by the data because the Coulomb peak at very small $t$ is sensitive to
the interference between the hadronic and Coulomb terms. 

However, this approach ignores information linking the phase of the amplitude to its 
variation with energy, so instead we assume that the hadronic amplitude is described by Regge theory and determine the unknown parameters in it from data beyond the Coulomb peak at a wide range of energies. The data in the Coulomb peak do not then play a part in determining the phase of the amplitude, but we do check that they are well described when the Coulomb
contribution is added to the amplitude.

Regge theory has long been known to give an excellent description of soft hadronic
processes\defref\collins{
P D B Collins, {\oblique Introduction to Regge Theory}, Cambridge
University Press (1977)
}\defref\book
{
A Donnachie, H G Dosch, P V Landshoff and O Nachtmann, {\oblique Pomeron Physics
and QCD}, Cambridge University Press (2002)
}. 
This paper applies it  in its simplest form to small-$t$ data from 13.76 GeV to 13 TeV for total cross sections 
and elastic scattering at small $t$, namely $|t|\leq 0.1$~GeV$^2$, by 
including in the amplitude the exchange of the soft pomeron $\P$, of the reggeons 
$\rho,\omega,f_2,a_2$ and of two pomerons $\P\P$.
The fit reveals no need\defref\nicolescu{
E Martynov and B Nicolescu, arXiv:1811.07635
}
for any odderon contribution at small $t$.

For the reggeon exchanges Regge theory introduces a trajectory
$\alpha(t)$ associated with Chew-Frautschi plots of the squares of masses of particles 
with the same quantum numbers but different spins.
According to figure 2.13 of reference {\book} the trajectories
are found to be exchange-degenerate, $\alpha_+(t)$ for $f_2,a_2$ and 
$\alpha_-(t)$ for $\rho,\omega$, with to a good approximation
$$
\alpha_{\pm}(t)=1+\epsilon_{\pm}+\alpha'_{\pm}t
\eqno(1a)
$$
where
$$
\epsilon_+=-0.3~~~\alpha'_+=0.8\hbox{ GeV}^{-2}~~~~~~~~~~~~
\epsilon_-=-0.56~~~\alpha'_-=0.92\hbox{ GeV}^{-2}
\eqno(1b)
$$
The trajectory for pomeron exchange is assumed similarly to be linear in $t$,
with intercept $1+\epsilon_{\P}$ and slope $\alpha'_{\P}$ determined by
the fit.

Each of these exchanges contributes
$$
X_iF(t)(2\nu\alpha'_i)^{\alpha_i(t)}\xi(t)~~~~~~~~i=\P,\pm
\eqno(1c)
$$
to the elastic amplitude, where
$$
2\nu=\half(s-u)=s-2m^2+\half t
\eqno(1d)
$$
and the signature factor 
$$
\xi(t)=-e^{-\half i\pi\alpha}\hbox{\ \    or\ \  }-ie^{-\half i\pi\alpha}
\eqno(1e)
$$
according to whether the $C$-parity of the exchange is even or odd. The signature factor
determines the complex phase of the contribution.  For each exchange there is
a real factor $X_iF(t)$ which is not determined by the theory. For simplicity we assume
the same function $F(t)$ for each
$$
F(t)=A e^{a_1t}+(1-A)e^{a_2t}
\eqno(1f)
$$

Regge theory has had many successes over more than half a century, with the above exchanges
giving the main contributions to a wide variety of reactions. But it is necessary also
to take account of the double-exchange contributions 
$\P\P,\P\rho,\P\omega,\dots$ to the amplitude, and perhaps 
triple or more. For simplicity we include only the first of these
$A_{\P\P}(s,t)$. The
trajectory for this exchange is known\ref{\book}:
$$
\alpha_{\P\P}(t)=1+2\epsilon_{\P}+\half\alpha'_{\P}t
\eqno(2a)
$$
However, while this determines the power of $s$ at large $s$, and
the complex phase of the term, that is all that is known about it.
To construct a simple model, we introduce the eikonal function
(see for example equation (2.49) of reference {\book})
$$
\chi(s,b)=-\log\Big(1+{i\over{8\pi^2s}}\int d^2q e^{i{\bf q.b}}A(s,-{\bf q}^2\Big)
\eqno(2b)
$$
so that
$$
A(s,-{\bf q}^2) =2is\int d^2b\, e^{-i{\bf q}.{\bf b}} ~(\chi-\half\chi^2+\dots )
\eqno(2c)
$$
A model that is sometimes used to calculate $A_{\P\P}(s,t)$
is to take $\chi(s,b)$ to include only the single-exchange contribution,
which would give, if we omitted the term $\half t$ in (1d) (which is justified 
because we are working only at small $t$)
$$ 
\tilde \chi_{\P}(s,b)={1\over{8i\pi^2s}}\int d^2qe^{i{\bf q.b}}A_{\P}(s,-{\bf q}^2)=
\sum_{i=1,2}{iZ_i\over{8\pi s D_i}}\exp\Big (\alpha_{\P}(0)L-{\bf b}^2/(4D_i)\Big )
\eqno(2d)
$$
with
$$
Z_1=X_{\P}A~~~~~~Z_2=X_{\P}(1-A)~~~~~~~~~
L=\log(2\nu\alpha_{\P}')-\half i\pi~~~~~~~~~D_i=a_i+\alpha_{\P}'L
\eqno(2e)
$$
Then the $\chi ^2$ term in (2c) would be the double-exchange contribution
to the amplitude,
$$
A_{\P\P}(s,t)=i\sum_{i,j=1,2}{Z_iZ_j\over{16\pi s(D_i+Dj)}}\exp\Big(2\alpha_{\P}(0)L+tD_iD_j/(D_i+D_j)\Big )
\eqno(2f)
$$
However, if instead we were to choose $\chi(s,b)$ to include also a
double-exchange contribution, when it is inserted into (2c) this would multiply (2f) by some constant $C$. 
Our fit finds that $C$ should be close to $\half$.

So we have 8 free parameters in addition to $C$, which we determine from the data beyond the Coulomb peak. We then add to the amplitude the photon-exchange term
$$
8\pi\alpha_{\hbox{\sevenrm EM}}G(t)/t
\eqno(2g)
$$
Here $G(t)$ is a squared form factor, equal to 1 at $t=0$. Choosing 
the square of either the Dirac or the Pauli form factor, or a combination 
of them,  gives almost the same result,
because the term is negligbly small except at extremely small $t$. 

The fits shown in figures 1 to 5 are with the choices
$$
\epsilon_{\P}=0.108~~~  X_{\P}= 166.3~~~  X_+= 201.2~~~  X_-= 119.8~~~ 
\alpha'_{\P}=0.321\hbox{ GeV}^{-2}~~~ 
$$$$
C=0.5~~~ A=0.561~~~  a_1=0.321\hbox{ GeV}^{-2}~~~ a_2= 7.674\hbox{ GeV}^{-2}
\eqno(3)
$$

As is seen in figure 5, the value 0.14 obtained for $\rho$ at 13 TeV is
rather different from that of at most 0.1 concluded by TOTEM from their 
data\defref{\totemrho}{
TOTEM collaboration: G Antchev et al, arXiv:1812.04732
}. The reason why the curves rise to a maximum and then fall again as the
energy increases is that the $\P\P$ term becomes progressively more important.

\bigskip
{\bold 2 Comments}

{\bold 1} With just 9 adjustable parameters, Regge theory provides a fit to data over a range of energies differing by a factor of 1000. The fit is extremely good, though less than perfect in some cases. There are also 
some anomalies in the data. An example is shown in figure 6: the data at 7, 8 and 13 TeV 
agree well with a single exponential in $t$, but the slope for the 7 TeV data lies
between that for 8 and 13 TeV, which is surely anomalous.

{\bold 2} In their extraction of $\rho$ from their 13~TeV data, TOTEM assume\ref{\totemrho}
that the ratio
of the real to the imaginary part of the hadronic amplitude is independent of $t$ from
0 to $-0.1$~GeV$^2$ or more. Figure 7 shows how it varies with t for the fit described
here.  

{\bold 3} TOTEM also extract $\rho$ by using the data only at the
one energy 13 TeV. 
If we perform our fit only at 13~TeV and include the West-Yennie
phase\defref\wy
{G B West and D R Yennie,Physical Review 172 (1968) 1413
} 
in the Coulomb term (2g) we obtain $\rho=0.1$, in agreement with TOTEM. 
However, it has been shown\defref\kl
{V Kundrat and M Lokajıcek, Physics Letters B656 (2007) 182
}
that this use of the West-Yennie phase is incorrect because of the 
variation of the amplitude's phase with $t$ shown in figure 7.

{\bold 4} Fitting the data at only one energy ignores information linking 
the phase of the amplitude to its variation with $s$: see the signature 
factors (1e) for example. If we include in our fit even only the 13~TeV
and 8~TeV data, the output for $\rho$ is 0.14 whether or not the
West-Yennie phase is included  though, as figure 8 shows, the fit is 
slightly better without it.

{\bold 5} It should be recognised that the value of $\rho$ extracted from data inevitably
depends on just what functional form is used to fit the data. This is illustrated
in figure 9, which shows that over a wide range of values of $\sqrt s$ the real
part of $\log(-s)$ agrees very well with a power of $s$, but the corresponding
imginary parts are somewhat different.

{\bold 6} Those who fit total cross section and elastic scattering data often 
replace
powers of $s$ by log factors. In Regge theory this is unnatural: it would correspond to
more than just a simple pole in the complex angular momentum plane. The excuse for
including a log, or the square of a log, in the amplitude is often said to be to to saturate the
Froissart-Lukaszuk-Martin bound\defref\flm
{
M~Froissart,  Physical Review {123} (1961) 1053;
L~Lukaszuk and A~Martin,  Il Nuovo Cimento { 47A} (1967) 265
}. 
However, the bound is about 20 barns at LHC energies and so is irrelevant.
\vskip 3pt
{\bold 3 Concluding remarks}

The value 0.14 obtained for $\rho$  does not encourage the belief\ref{\nicolescu}
that there is
an odderon contribution at $t=0$. However, there is good reason to believe that there
is an odderon contribution at large $t$ and that it is identified with triple-gluon
exchange. Indeed, this led us to predict\defref\predict{
A Donnachie and P V Landshoff, Physics Letters 123B (1983) 345
}
that $pp$ and $\bar pp$ scattering would be different, as was confirmed\defref\breakstone{
A Breakstone et al, Physical Review Letters 54 (2180) 1985
}
at the CERN ISR. We included such a term in a previous fit\defref\elastic{
A Donnachie and P V Landshoff, Physics Letters B727 (2013) 500
}. Note, however, that to lowest order in the strong coupling triple-gluon 
exchange's contribution to the $pp$ amplitude is real positive, while
of the TOTEM odderon would have negative real part at $t=0$.
\vskip 3pt
We are grateful to Professor J R Cudell for his comments in early stages of this work.
\medskip\immediate\closeout\rfile\writestoppt
\baselineskip=0pt{{\bf References}}\h{\frenchspacing%
\parindent=20pt\escapechar=` \input refs.tmp\bigskip}\nonfrenchspacing

\pageinsert{
\epsfxsize=0.45\hsize\epsfbox[60 60 320 295]{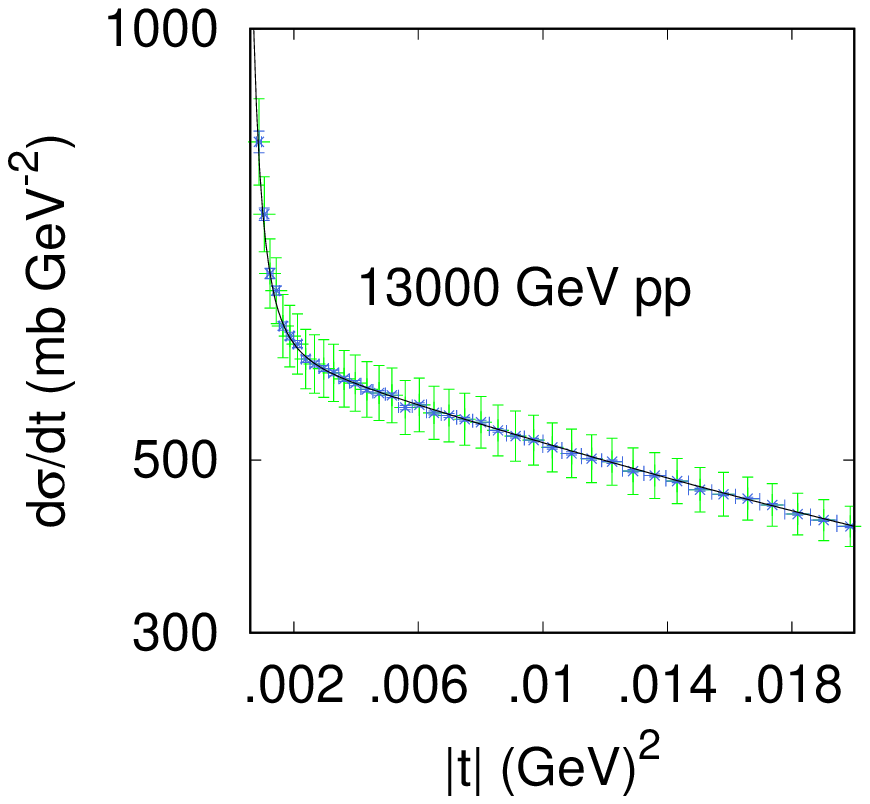}
\hfill
\epsfxsize=0.45\hsize\epsfbox[60 60 320 295]{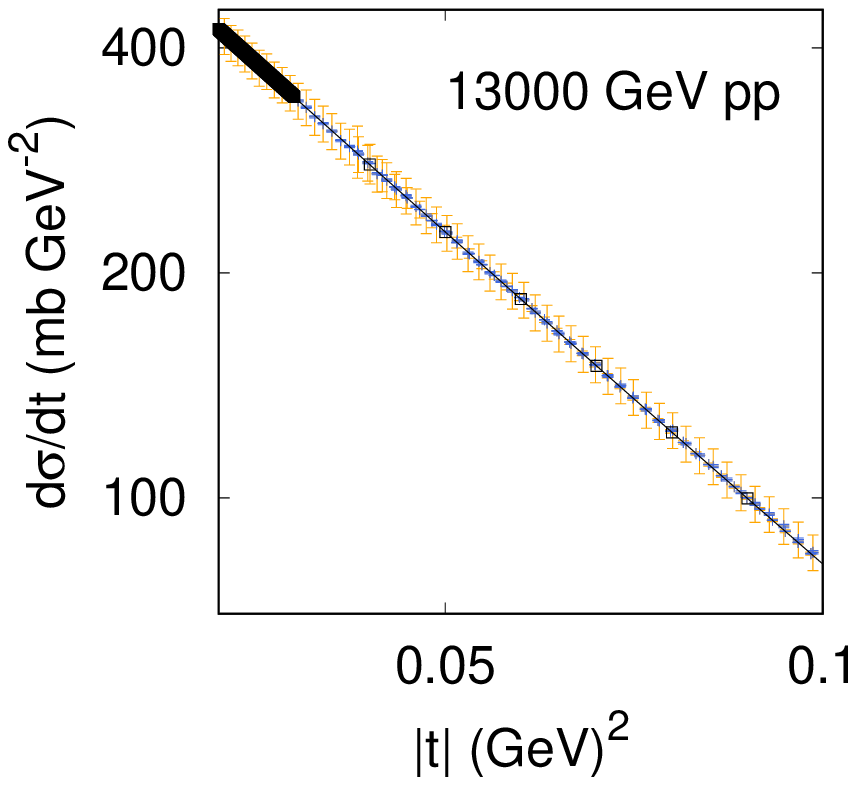}
\bigskip
\epsfxsize=0.45\hsize\epsfbox[60 60 320 295]{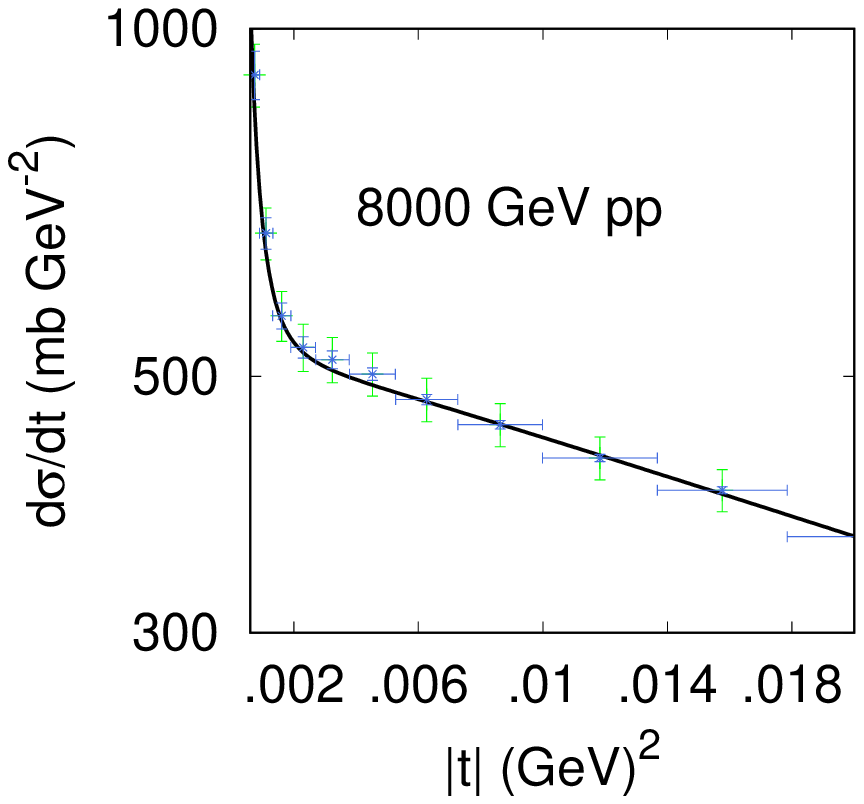}
\hfill
\epsfxsize=0.45\hsize\epsfbox[60 60 320 295]{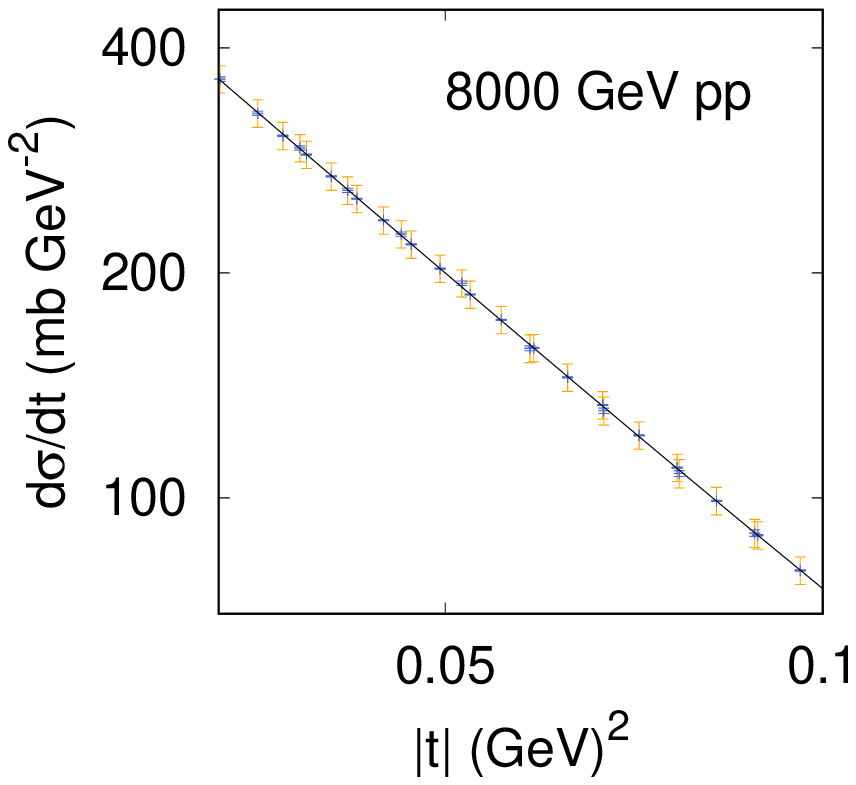}
\bigskip
\epsfxsize=0.45\hsize\epsfbox[60 60 320 295]{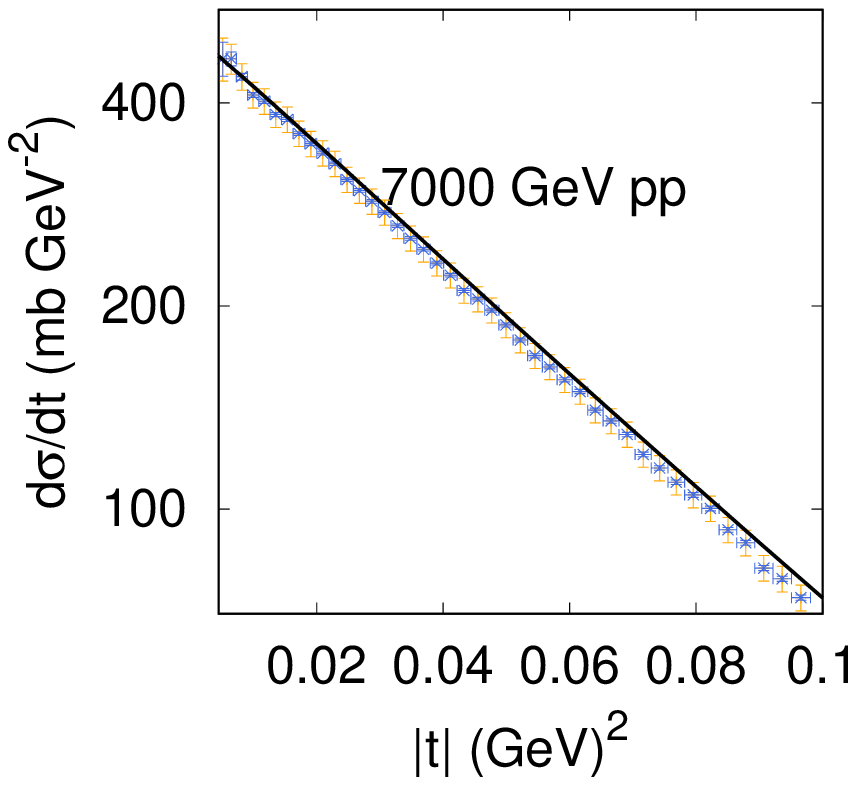}
\hfill
\epsfxsize=0.45\hsize\epsfbox[60 60 320 295]{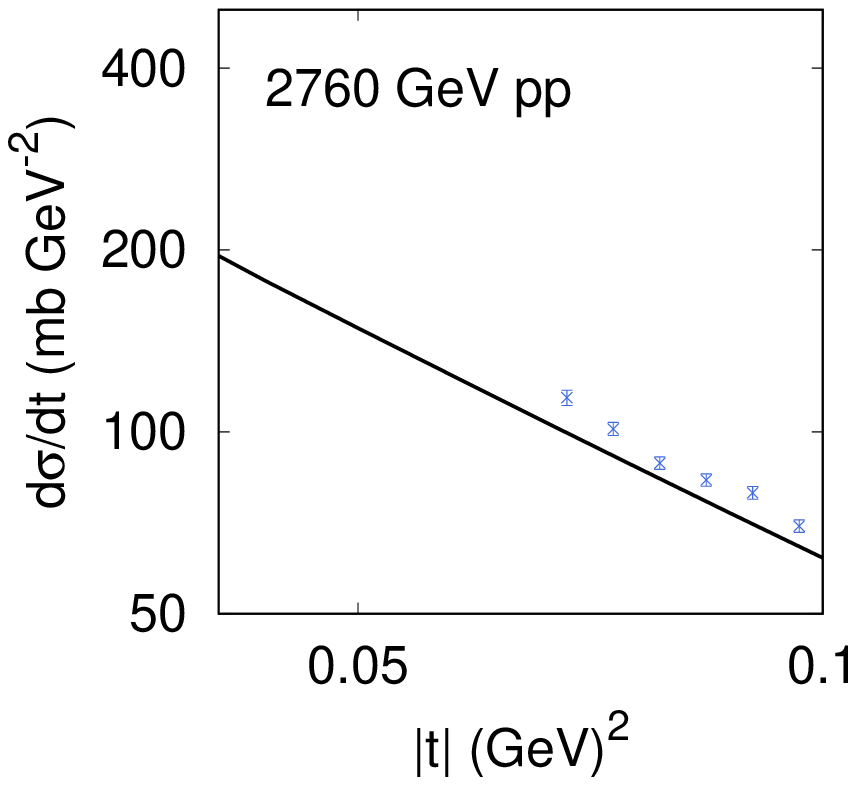}
\bigskip
Figure 1: Fits to TOTEM data\defref\totem{
TOTEM collaboration: G Antchev et al, European Physics Letters 101 (2013) 21002,    
arXiv:1610.0060, arXiv:1812.08283, arXiv:1812.08610
}\ref{\totemrho}

}
\endinsert
\vfill\eject
\pageinsert{
\epsfxsize=0.45\hsize\epsfbox[60 60 320 295]{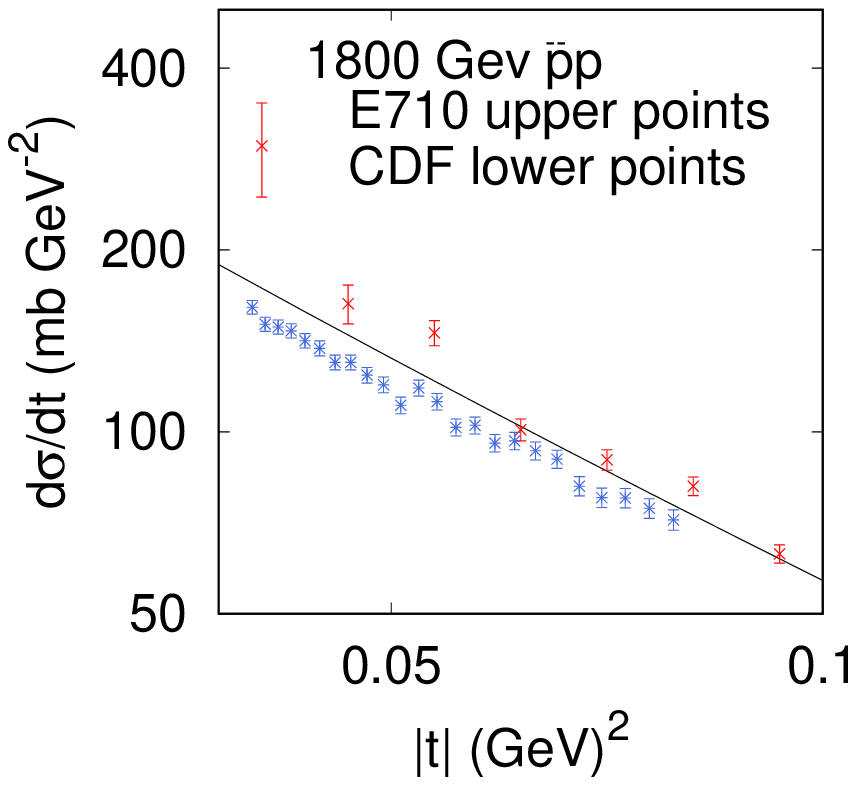}
\hfill
\epsfxsize=0.45\hsize\epsfbox[60 60 320 295]{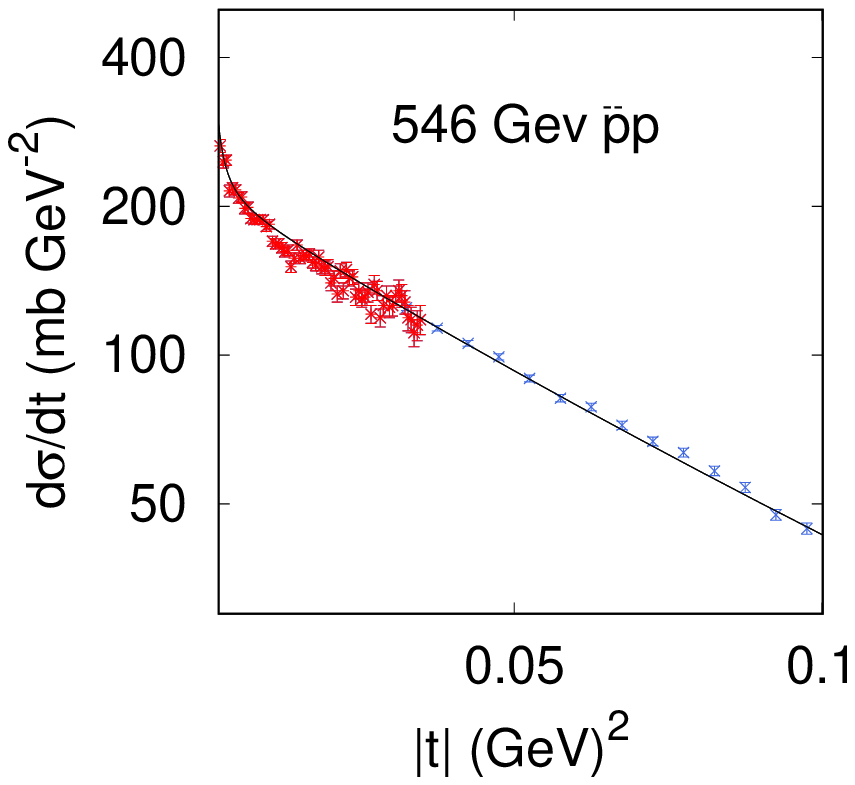}
\bigskip
Figure 2: Fits to CERN and Fermilab $p\bar p$ data (which are referenced
in reference {\elastic})
\bigskip
\epsfxsize=0.45\hsize\epsfbox[60 60 320 295]{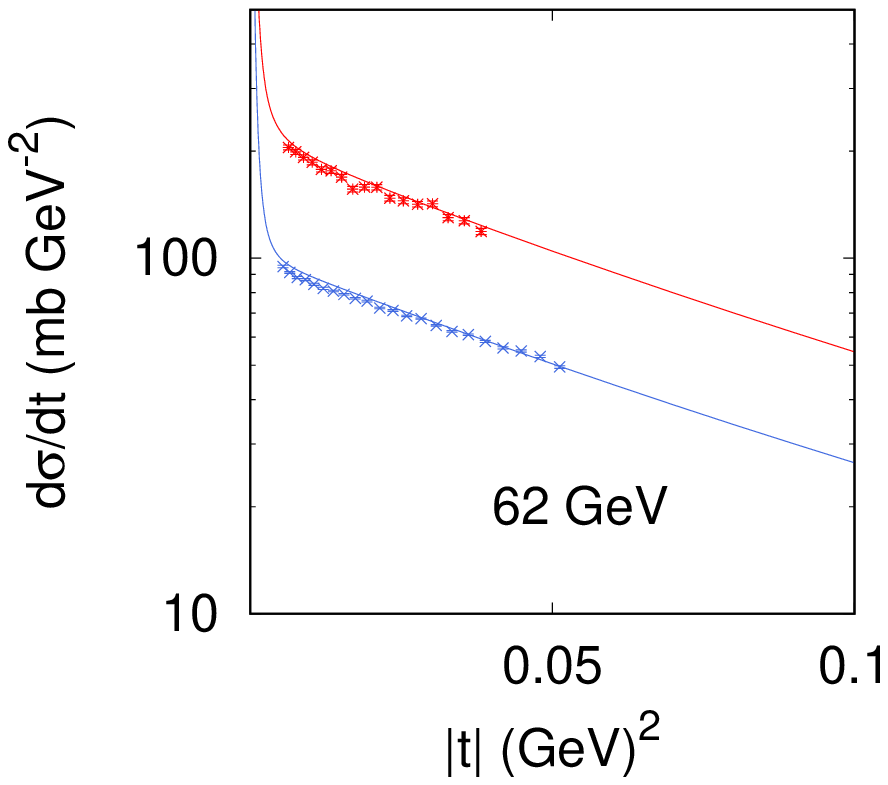}
\hfill
\epsfxsize=0.45\hsize\epsfbox[60 60 320 295]{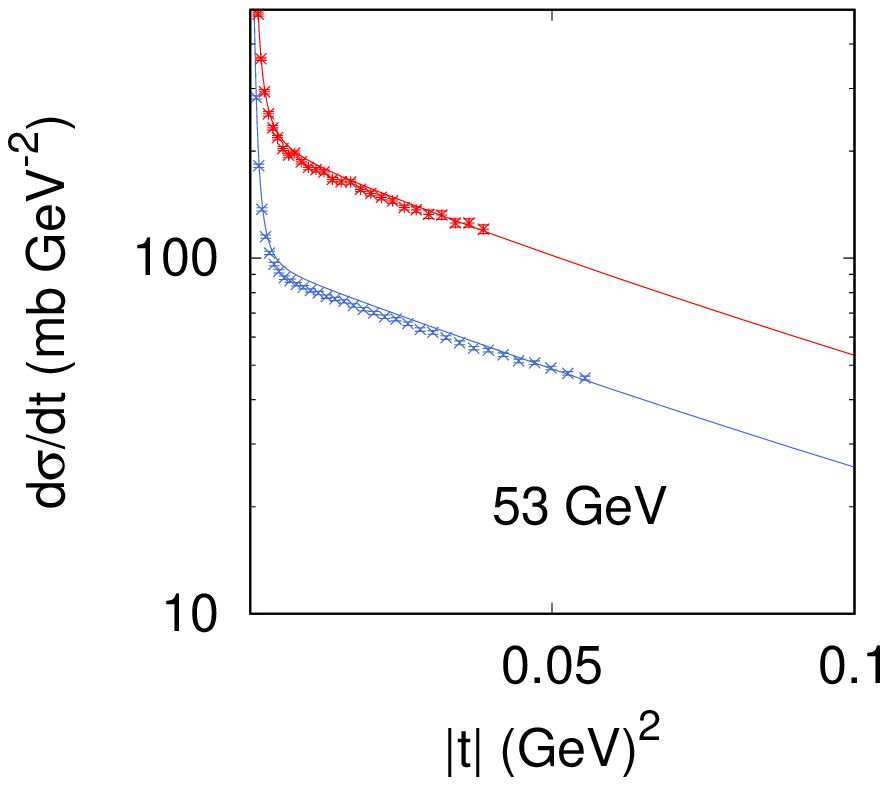}
\bigskip
\epsfxsize=0.45\hsize\epsfbox[60 60 320 295]{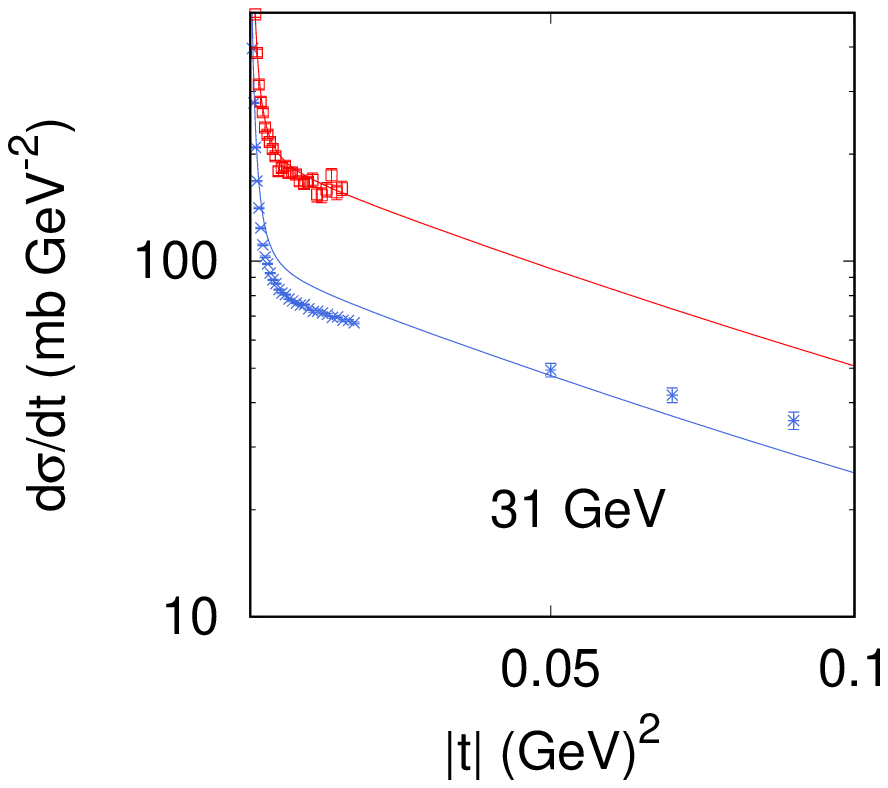}
\hfill
\epsfxsize=0.45\hsize\epsfbox[60 60 320 295]{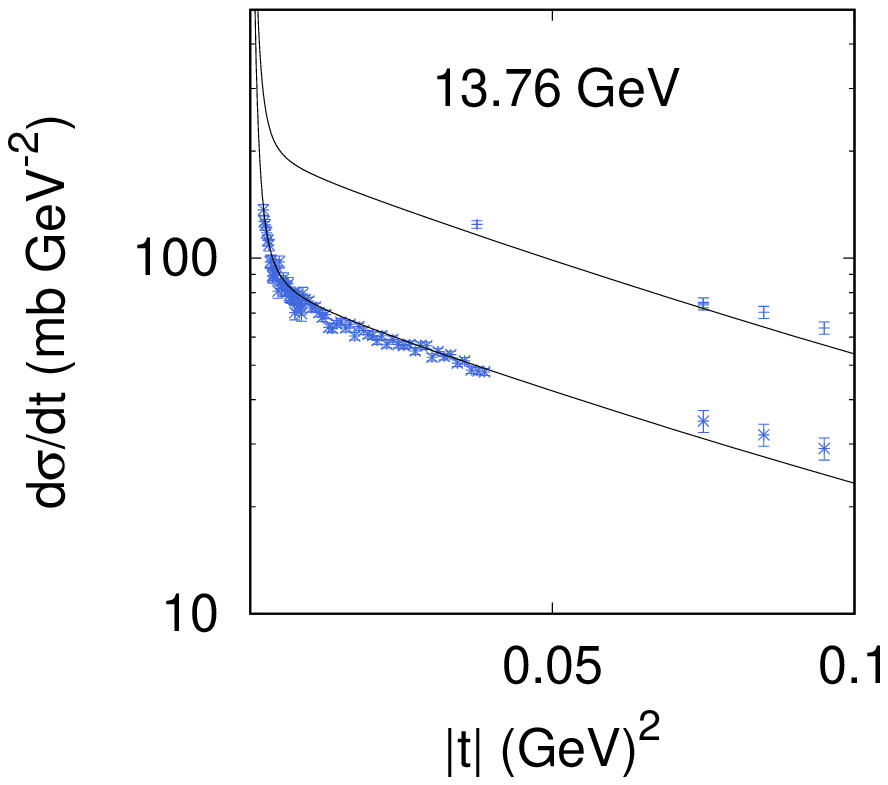}
\bigskip
Figure 3: Fits to fixed target and CERN ISR data (which are referenced
in reference {\elastic}). The lower points are $pp$ scattering, the upper points $p\bar p$ multiplied by 2.
}
\endinsert
\vfill\eject
\centerline{
\epsfxsize=0.45\hsize\epsfbox[60 60 390 295]{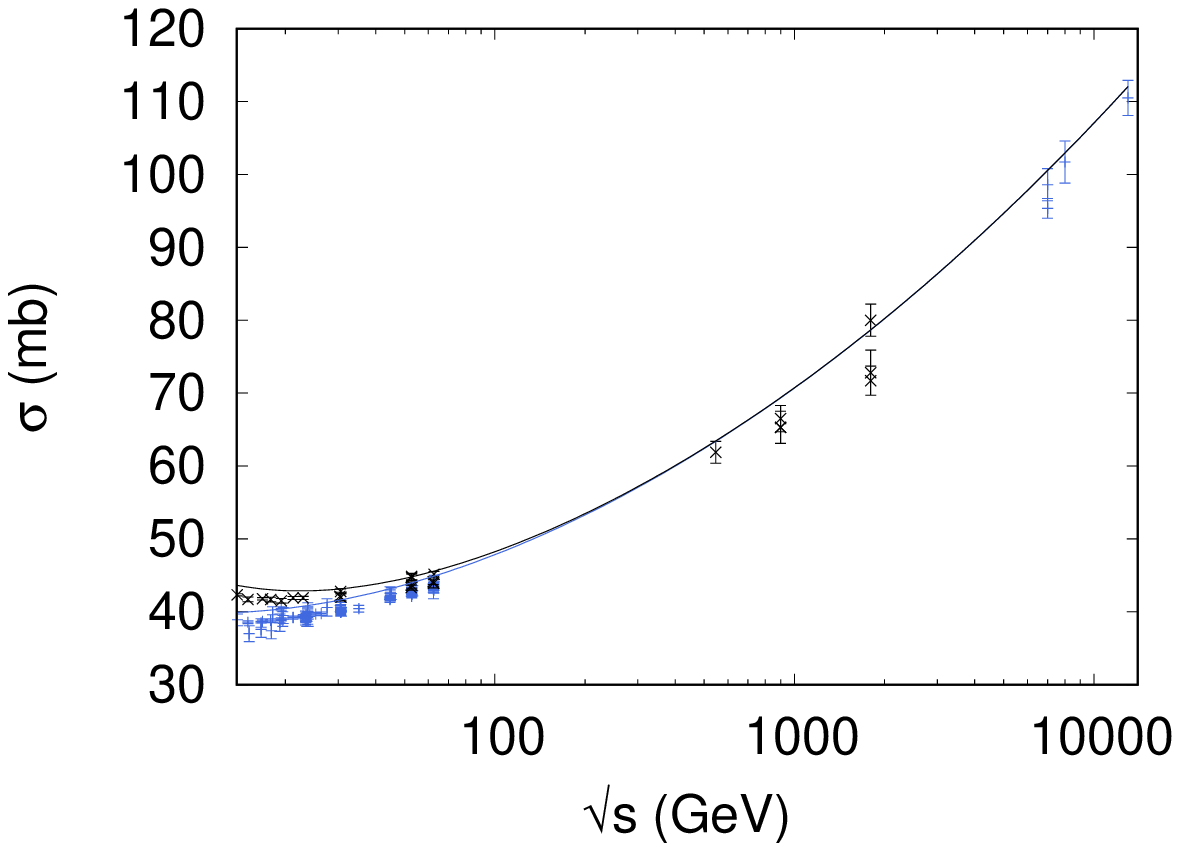}\phantom{XXXXX}}
\bigskip
Figure 4: Fits to the $pp$ and $p\bar p$ total cross sections
\bigskip\bigskip
\epsfxsize=0.45\hsize\epsfbox[90 60 370 295]{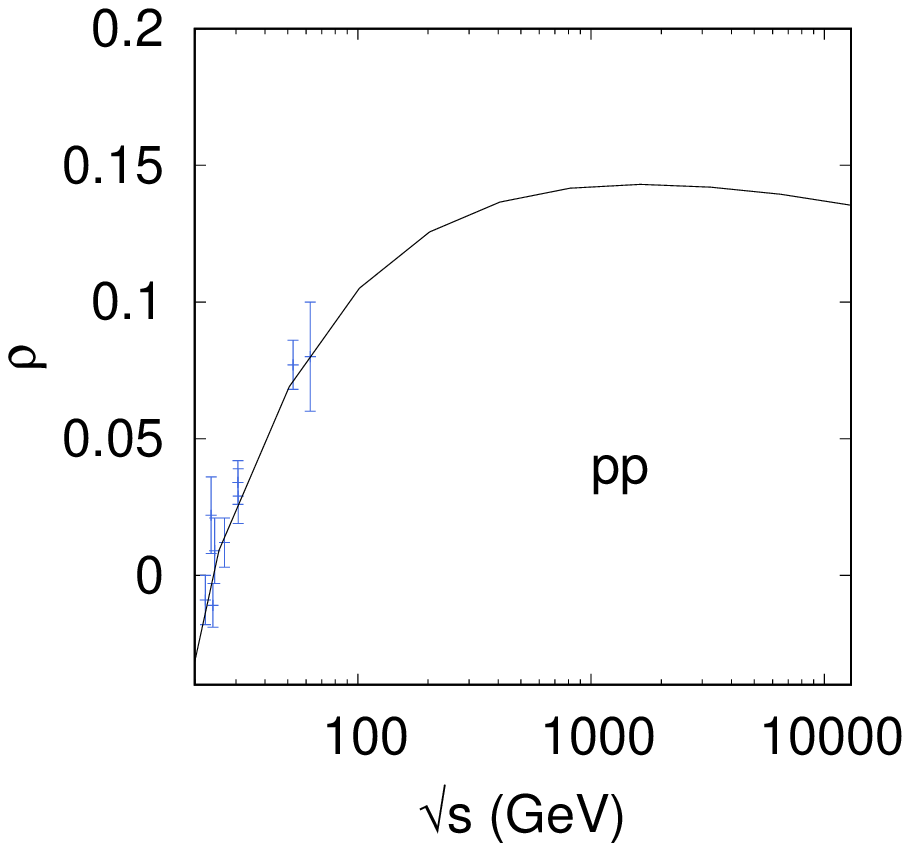}
\hfill
\epsfxsize=0.45\hsize\epsfbox[90 60 370 295]{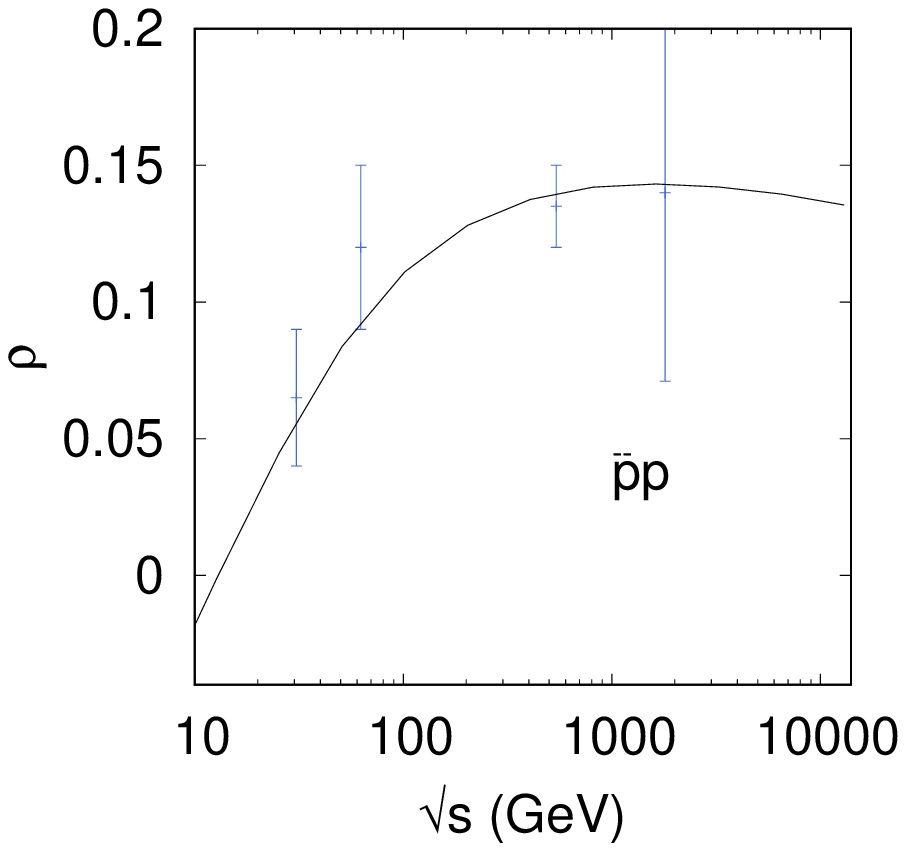}
\bigskip
Figure 5: Outputs for $\rho$
\bigskip\bigskip
\centerline{
\epsfxsize=0.5\hsize\epsfbox[60 60 390 295]{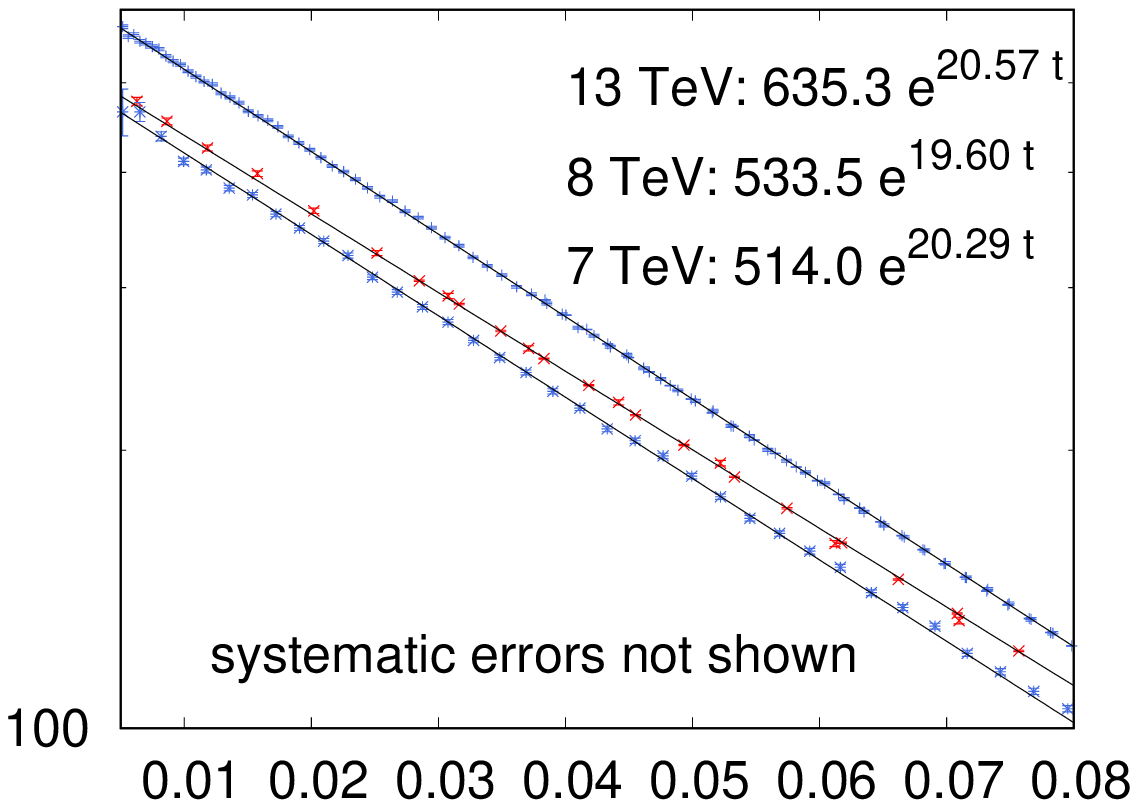}
}
\bigskip
Figure 6: Exponential fits to TOTEM data
\bigskip
\bigskip
\centerline{
\epsfxsize=0.45\hsize\epsfbox[60 60 390 295]{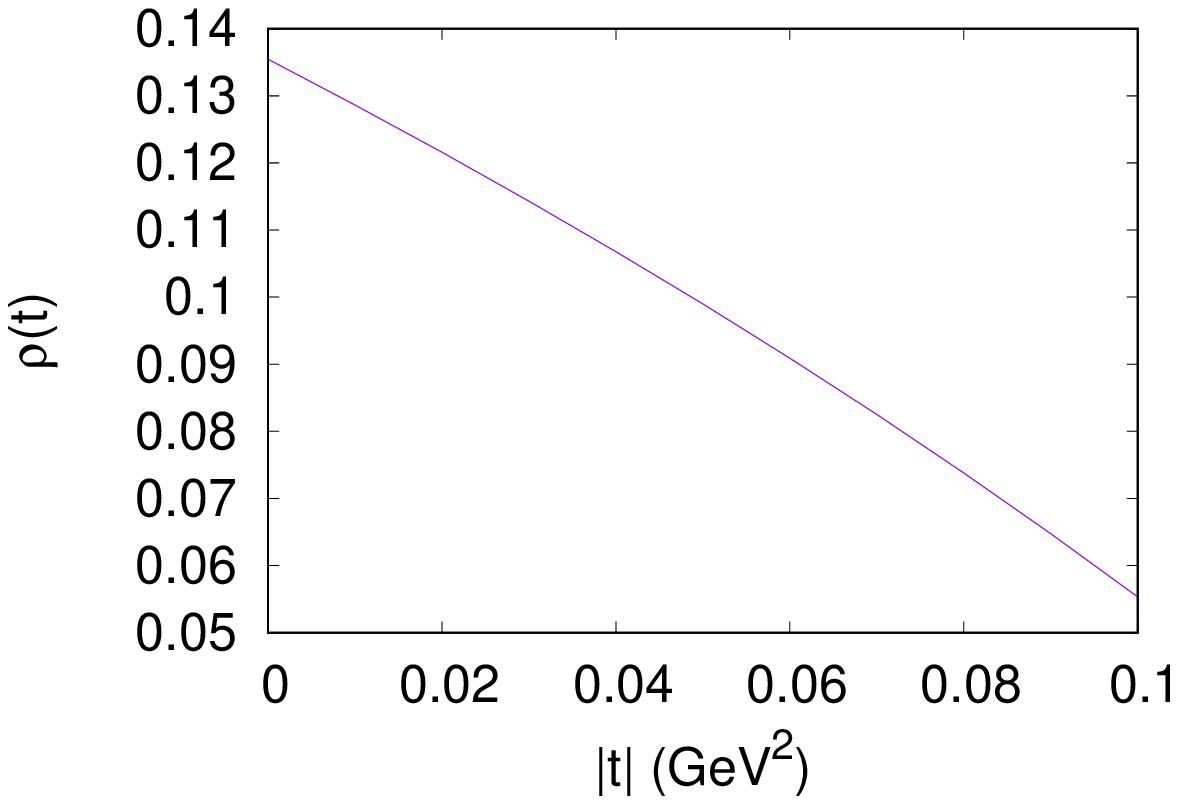}
}
\bigskip
Figure 7: The calculated ratio $\rho(t)$ of the real to the imaginary part of the hadronic 
amplitude at 13~TeV.
\bigskip\bigskip
\centerline{
\epsfxsize=0.5\hsize\epsfbox[60 60 390 295]{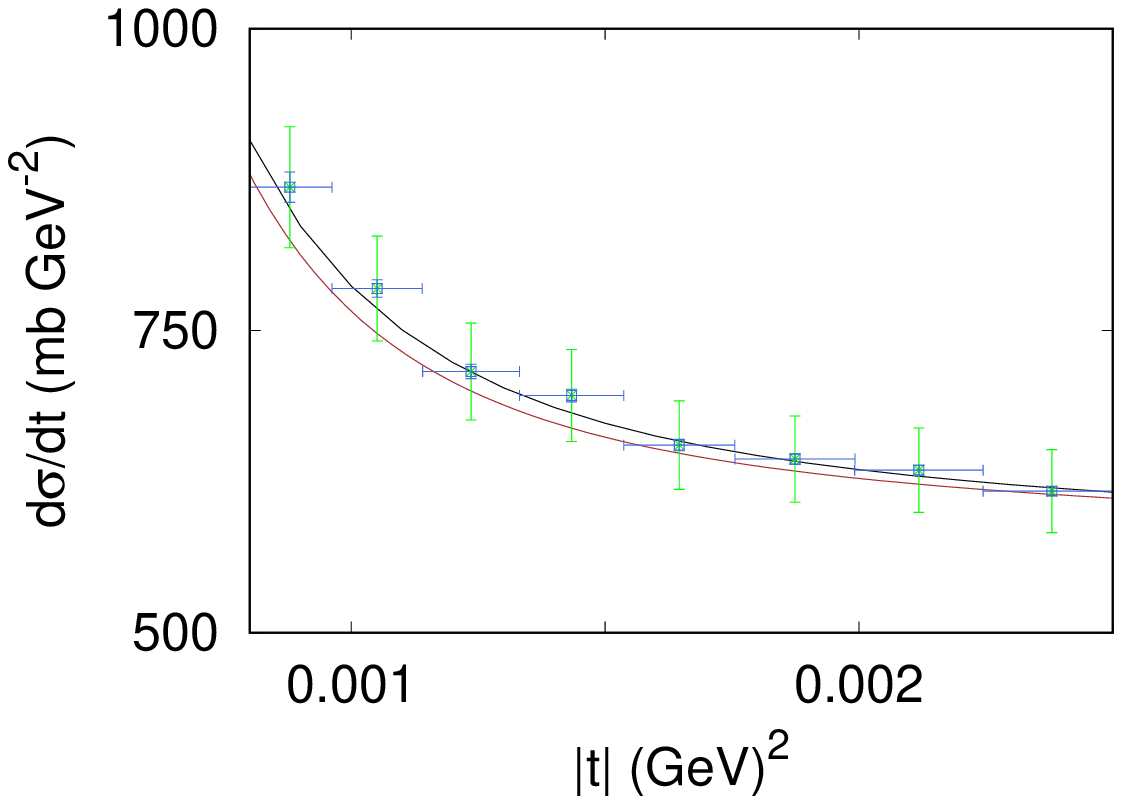}
}
\bigskip
Figure 8: 13 TeV data without using West-Yennie phase (upper curve) and with
(lower curve)
\bigskip\bigskip
\centerline{
\epsfxsize=0.45\hsize\epsfbox[60 60 390 295]{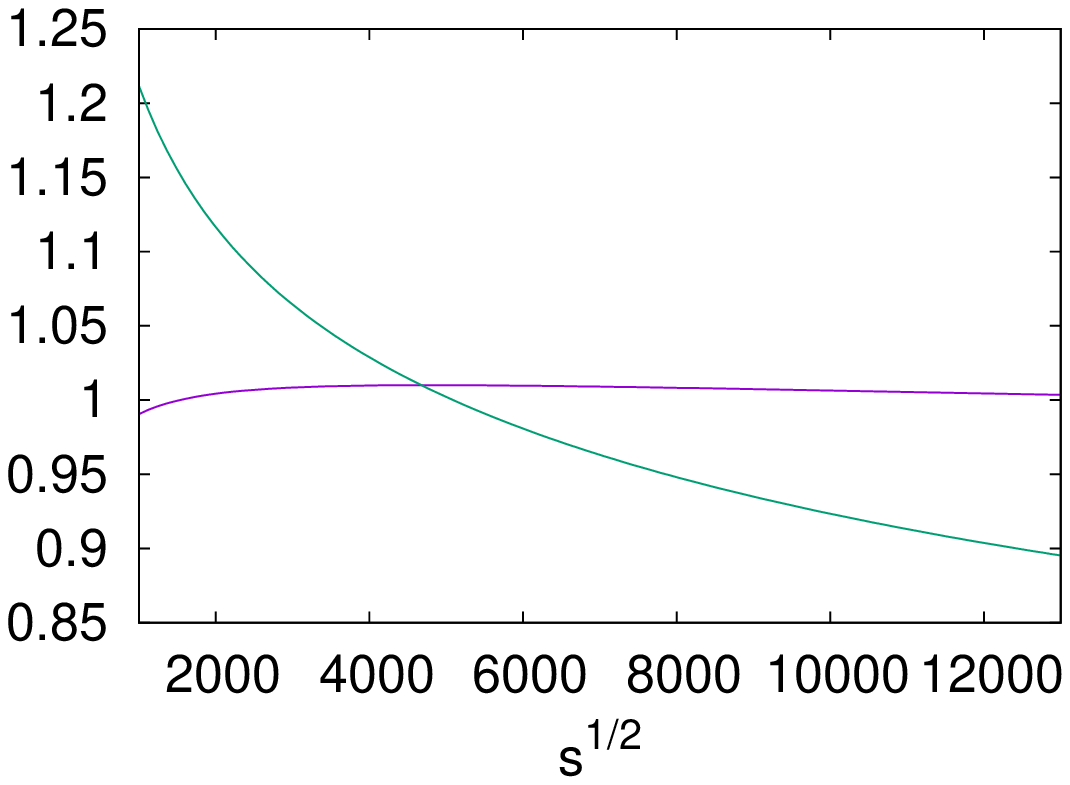}
}
\bigskip
Figure 9: Ratios of real and imaginary parts of $\log (-s)$ to those of 
$Cs^{\epsilon}e^{-i\pi\epsilon/2}$ with $C=6.2$ and $\epsilon=0.059$,
chosen so that the real parts almost agree  
\bye